\documentclass[reprint,superscriptaddress,longbibliography,eqsecnum,nofootinbib,amsmath,amssymb,aps,floatfix,onecolumn]{revtex4-2}

\newcommand\be{\begin{equation}}
\newcommand\ee{\end{equation}}
\newcommand\nono{\nonumber}
\newcommand\bse{\begin{subequations}}
\newcommand\ese{\end{subequations}}

\usepackage[utf8]{inputenc}
\usepackage[T1]{fontenc}
\usepackage[dvipsnames]{xcolor}
\usepackage{graphicx}
\usepackage{bm}
\usepackage[colorlinks=true,allcolors=blue]{hyperref}
\usepackage{mathrsfs}
\usepackage[utf8]{inputenc}
\usepackage{leftidx}

\begin{document}
\title{Singular Lagrangians, Constrained Hamiltonian Systems and\\ Gauge Invariance: An Example of the Dirac--Bergmann Algorithm}
\author{J. David Brown}
\email{david\_brown@ncsu.edu}
\affiliation{Department of Physics, North Carolina State University, Raleigh, NC 27695}
\date{\today}

\begin{abstract} 
The Dirac--Bergmann algorithm is a recipe for converting a theory with a singular Lagrangian into a constrained Hamiltonian system. Constrained Hamiltonian systems include gauge theories---general relativity, electromagnetism, Yang--Mills, string theory, etc. The Dirac--Bergmann algorithm is elegant but at the same time rather complicated. It consists of a large number of logical steps linked together by a subtle chain of reasoning. Examples of the Dirac--Bergmann algorithm found in the literature are designed to isolate and illustrate just one or two of those logical steps. 
In this paper,
 I analyze a finite-dimensional system that exhibits all of the major steps in the algorithm. The system includes primary and secondary constraints, first and second class constraints, restrictions on Lagrange multipliers, and both physical and gauge degrees of freedom. This relatively simple system provides a platform for discussing the Dirac conjecture, constructing Dirac brackets, and applying gauge conditions.
\end{abstract}
\maketitle

\section{Introduction}\label{sec:introduction}
In the early 1950's,
 Dirac and Bergmann  independently developed the Hamiltonian formalism for systems with singular Lagrangians~\cite{Bergmann1949,BergmannBrunings1949,Dirac1950,BergmannEtAl1950,Dirac1951,AndersonBergmann1951,BergmannGoldberg1955,Dirac1958a,Dirac1958b,DiracLectures}.
These systems,  often called ``constrained Hamiltonian systems'', include gauge theories. Gauge freedom is more clearly and more completely displayed in the Hamiltonian setting, with~the generators of gauge transformations expressed 
as functions on phase space. Historically, the~main motivation for casting gauge theories in Hamiltonian form was to facilitate their canonical quantization. Dirac and Bergmann were primarily motivated by the prospect of developing a quantum theory of gravity based on a Hamiltonian formulation of general~relativity. 

Textbook treatments of Lagrangian and Hamiltonian mechanics invariably assume that the Lagrangian $L(q,\dot q)$ is nonsingular; that is, 
that the matrix of second derivatives of $L(q,\dot q)$ with respect to the velocities is invertible. In~classical mechanics, 
the nonsingular case appears to be sufficient to 
cover problems of physical interest. However,  one might argue that textbooks avoid certain physically 
interesting 
problems simply because their Lagrangians are~singular. 

In field theory, 
the issue of singular Lagrangians and gauge freedom cannot be avoided. Nearly every 
field theory of physical interest---electrodynamics, Yang--Mills theory, general relativity, relativistic string theory---has gauge~freedom.  

The Dirac--Bergmann algorithm transforms a singular Lagrangian system into a Hamiltonian system. The~formalism is elegant but at the same time rather 
complex. It consists of a large number of logical steps, linked together by a chain of reasoning that can be difficult to keep straight.
 Of~course,
  there are many examples in the literature in which the Dirac--Bergmann algorithm is applied, converting a singular Lagrangian into Hamiltonian form. 
  However,~to my knowledge, all of these examples are designed to illustrate just one or two of the logical steps in the algorithm. The~student of the subject is faced with the task of linking  these examples together to create a complete picture of the~algorithm. 

For those who learn by example, what is needed is a single example that illustrates all of the major logical steps in the Dirac--Bergmann algorithm and shows how these steps are linked together. 
Such a ``complete'' example is not easy to identify because there is no obvious way to predict, starting with a particular Lagrangian, which of the steps in the algorithm will be~needed. 

The system analyzed in this paper is defined by the Lagrangian 
\be\label{TheLagrangian}
    L(q,\dot q) = \frac{1}{2} \Bigl\{(q_1 + \dot q_2 + \dot q_3 )^2 + (\dot q_4 - \dot q_2)^2 
    + (q_1 + 2 q_2)(q_1 + 2q_4) \Bigr\}
\ee
where the dot denotes a time derivative.  The matrix of second derivatives  with respect to the velocities is
\be
    \frac{\partial^2 L}{\partial \dot q_i \partial \dot q_j} = \begin{pmatrix} 
        0 & 0 & 0 & 0 \\ 0 & 2 & 1 & -1\\ 0 & 1 & 1 & 0 \\ 0 & -1 & 0 & 1 \end{pmatrix} \ .
\ee
This matrix is singular; it has rank~2.

As we will see, the~system defined by the Lagrangian (\ref{TheLagrangian}) is relatively complete.\footnote{This example does not cover every contingency. In~particular, it does not include redundant constraints~\cite{HenneauxTeitelboim}.} It contains both primary and secondary constraints, both first and second class constraints, and~restrictions on the Lagrange multipliers. The~first class constraints for this system are not all primary; this allows us to address the Dirac conjecture. The~second class constraints can be eliminated by introducing Dirac brackets. Finally, this system contains both physical and gauge degrees of freedom. The~gauge freedom can be eliminated with suitable gauge~conditions.  

One characteristic of any complete example such as  (\ref{TheLagrangian}) is that the configuration space, the~space of $q$'s, must be at least four-dimensional. 
Here is why: 
The number of physical degrees of freedom is equal to the dimension of the configuration space, minus the number of first class constraints, minus half the number of second class constraints. 
If the example is to have at least one physical degree of freedom, at~least two first class constraints (one primary and one secondary), and~at least two second class constraints (the number of second class constraints must be even), then the configuration space must be at least~four--dimensional.  

The study of constrained Hamiltonian systems  predates Dirac and Bergmann with 
earlier work by Rosenfeld~\cite{RosenfeldGerman,RosenfeldEnglish}. Like Rosenfeld, Bergmann and his collaborators~\mbox{\cite{Bergmann1949,BergmannBrunings1949,BergmannEtAl1950,AndersonBergmann1951,BergmannGoldberg1955,Penfeld1951}} were focused on field theories such as general relativity that are covariant with respect to 
general four-dimensional coordinate 
transformations.  (For a historical review, see \cite{Salisbury:2006dul}.) 
Dirac took a more basic approach to the problem by considering a generic singular Lagrangian~\cite{Dirac1950,Dirac1951,Dirac1958a,Dirac1958b,DiracLectures}. He developed the algorithm for the case of systems with a finite number of degrees of freedom. His view was that the generalization to field theory, with~an infinite number of degrees of freedom, would be ``merely a formal~matter''.  

Although the typical starting point for the Dirac--Bergmann analysis is a singular Lagrangian, 
not all gauge theories can be expressed in terms of a Lagrangian that depends only on the $q$'s. General relativity is one such example. The~Einstein--Hilbert action is a functional of Lagrange multipliers (the lapse function and shift vector) as well as the configuration variables (the spatial metric). Nevertheless, the~Dirac--Bergmann algorithm provides the foundation for our interpretation of general relativity as a constrained Hamiltonian~system.

In this paper,
 I apply the Dirac--Bergmann algorithm to the Lagrangian (\ref{TheLagrangian}), following closely the general treatment given by Henneaux and Teitelboim~\cite{HenneauxTeitelboim}. In~turn, the~account of Henneaux and Teitelboim closely follows Dirac's 1964 {\em Lectures on Quantum Mechanics}~\cite{DiracLectures}. Presentations of the Dirac--Bergmann algorithm can also be found in books by 
Hanson, Regge and Teitelboim~\cite{HansonReggeTeitelboim}, Sundermeyer~\cite{Sundermeyer},   Rothe and Rothe~\cite{RotheRothe}, and~Lusanna~\cite{Lusanna:2019cmq}.  

Throughout the paper,
 I attempt to explain the reasoning behind the logical steps of the Dirac--Bergmann algorithm, but~avoid general proofs. The~reader is referred to references~\cite{HansonReggeTeitelboim,Sundermeyer,HenneauxTeitelboim,RotheRothe,Lusanna:2019cmq} for more~details. 

We begin in Section~\ref{sec:lagrangianalysis} with a derivation of Lagrange's equations for the singular Lagrangian (\ref{TheLagrangian}). The~general solution is derived, and~in Section~\ref{sec:gaugeinvariance}, we discuss the gauge freedom at the Lagrangian level.  We begin construction of the Hamiltonian theory in Section~\ref{sec:primaryconstraints} with a derivation of the  primary constraints and canonical Hamiltonian. In~Section~\ref{sec:primaryhamiltonian},
 we introduce the primary Hamiltonian and the primary action.  Section~\ref{sec:hamiltonsequations} is devoted to a discussion of the initial value problem and the need to go beyond the primary Hamiltonian.
 In~Section~\ref{sec:consistencyconditions},
  we apply Dirac's consistency conditions to derive the secondary constraints and restrictions on the Lagrange multipliers. The~concept of weak equality is introduced in Section~\ref{sec:weakequality}, along with a formal analysis of the restrictions on the Lagrange multipliers. The~total Hamiltonian is computed in Section~\ref{sec:totalhamiltonian}, and~in Section~\ref{sec:firstandsecondclass} 
   we sort the constraints into first and second class. The~first class Hamiltonian and gauge generators are identified in Section~\ref{sec:firstclasshamiltonian}, where we also introduce the Dirac conjecture. In~Sections~\ref{sec:extendedH} and \ref{sec:extendedS},
    we define the extended Hamiltonian and extended action. Dirac brackets are used in Section~\ref{ref:diracbrackets} to eliminate the second class constraints, which yields a partially reduced Hamiltonian. The~corresponding partially reduced action is derived in Section~\ref{sec:partiallyreducedS}, and~in Section~\ref{sec:partiallyreducedL} 
     we eliminate the momenta to obtain a partially reduced Lagrangian.
       Gauge conditions are introduced in Section~\ref{sec:gaugeconditions} and Dirac brackets are used to eliminate the constraints and gauge conditions. This yields a  fully reduced Hamiltonian. The~fully reduced action is derived in Section~\ref{sec:fullyreducedS}. Finally, Section~\ref{sec:summary} contains a summary of the Dirac--Bergmann algorithm and a discussion 
of the Einstein--Hilbert action for general~relativity.

\section{Lagrangian~Analysis}\label{sec:lagrangianalysis}
The action is the integral of the Lagrangian (\ref{TheLagrangian}): 
\be
    S[q] = \int_{0}^{T} dt \, L(q,\dot q) \ .
\ee
The notation $S[q]$ indicates that $S$ is a functional 
of the complete set of coordinates, $q_i = \{ q_1,q_2,q_3,q_4\}$.  
The equations of motion are obtained by 
extremizing the action. For~this example, 
we are not concerned with boundary conditions and integrate by parts freely. 
Lagrange's equations are 
\bse\label{LagrangesEqns} 
\begin{align}
    0 &=  \frac{\delta S}{\delta q_1} = \dot q_2 + \dot q_3 + 2q_1 + q_2 + q_4 \ , \\
    0 &=  \frac{\delta S}{\delta q_2} =  - 2\ddot q_2 - \ddot q_3 + \ddot q_4 - \dot q_1 + q_1 + 2q_4 \ , \\
    0 &=  \frac{\delta S}{\delta q_3} =  - \ddot q_2 - \ddot q_3 -\dot q_1 \ , \\
    0 &=  \frac{\delta S}{\delta q_4} = \ddot q_2 -\ddot q_4 + q_1 + 2q_2 \ .
\end{align}
\ese
We can rewrite these as follows. First, add Equations~(\ref{LagrangesEqns}b) and (\ref{LagrangesEqns}d), then 
subtract Equation~(\ref{LagrangesEqns}c). This gives 
\bse\label{Leqn}
\be
    q_1 + q_2 + q_4 = 0 \ .
\ee
Next, subtract this result from Equation~(\ref{LagrangesEqns}a) to obtain 
\be
   \dot q_2 +\dot q_3 + q_1 = 0 \ .
\ee
The time derivative of this equation yields Equation~(\ref{LagrangesEqns}c).  Finally, we find the result
\be
    \ddot q_4 - \ddot q_2 = q_2 - q_4 \ .
\ee
\ese
by solving Equation~(\ref{Leqn}a) for $q_1$ and using the equation of motion (\ref{LagrangesEqns}d).

Equations~(\ref{Leqn}) are equivalent to 
Lagrange's Equations~(\ref{LagrangesEqns}). In~particular, Equation~(\ref{LagrangesEqns}a) is the sum of Equations~(\ref{Leqn}a) and (\ref{Leqn}b); Equation~(\ref{LagrangesEqns}b) is the sum of Equations~(\ref{Leqn}a) and (\ref{Leqn}c) with the 
time derivative of (\ref{Leqn}b) subtracted; Equation~(\ref{LagrangesEqns}c) is the negative of 
the time derivative of (\ref{Leqn}b); Equation~(\ref{LagrangesEqns}d) is obtained by 
subtracting (\ref{Leqn}c) from (\ref{Leqn}a). 

The equations of motion for this simple linear system are easily solved. Note that the combination $q_4 - q_2$ is determined by (\ref{Leqn}c) along with  initial or boundary data;  thus, we have 
\be
    q_4(t) - q_2(t) = A\sin t + B\cos t \ ,
\ee
where $A$ and $B$ are constants. 
Now,
 Equation~(\ref{Leqn}a) gives 
\be
    q_1(t) + 2q_2(t) = -A\sin t - B\cos t \ .
\ee
If we knew $q_2(t)$, we could solve the previous two equations  for $q_1(t)$ and $q_4(t)$, then integrate Equation~(\ref{Leqn}b) to obtain $q_3(t)$. 
Clearly,
 we do not have enough information to fully determine each of the $q$'s as functions of time. 
  One of the 
$q$'s must remain undetermined. For~example, let us  choose $q_2$ arbitrarily by setting $q_2(t) = -\Psi(t)/2$ for some function $\Psi(t)$. We can then use the equations above to solve for $q_1$, $q_3$ and $q_4$:
\bse\label{LagEqSolutions}\begin{align}
    q_1(t) & =  -A\sin t - B\cos t +\Psi(t) \ , \\
    q_2(t) & =  -\Psi(t)/2 \ ,\\
    q_3(t) & =  -A\cos t + B\sin t + \Psi(t)/2  - \int_{0}^t ds\, \Psi(s) + C \ ,\\
    q_4(t) & =  A\sin t + B\cos t - \Psi(t)/2 \ ,
\end{align}\ese
where $C$ is an integration constant. 
This is the general solution of the equations of~motion. 

\section{Gauge~Invariance}\label{sec:gaugeinvariance}
The undetermined function $\Psi(t)$ that appears in the general solution (\ref{LagEqSolutions}) can be freely specified. This is the gauge freedom of the theory. 
We can express the gauge freedom in another way: the Lagrangian (\ref{TheLagrangian}) and 
the equations of motion (either (\ref{LagrangesEqns}) or (\ref{Leqn})) are invariant under the replacements 
\bse\label{gaugewithepsilon}\begin{align}
    q_1(t) & \to  q_1(t) + \Psi(t) \ ,\\
    q_2(t) & \to  q_2(t) - \Psi(t)/2 \ ,\\
    q_3(t) & \to  q_3(t) + \Psi(t)/2 - \int_{0}^t ds\,\Psi(s) \ ,\\
    q_4(t) & \to  q_4(t) - \Psi(t)/2 \ ,
\end{align}\ese
where $\Psi(t)$ is an arbitrary function of~time. 

Although each configuration of the system (that is, each set of $q$ values)  corresponds to a specific physical state of the system, the~converse is not true. Because~of the gauge freedom, there are many sets of $q$'s that describe one and the same physical~state. 

Let us examine the gauge freedom more closely, in~anticipation of the Hamiltonian description of  evolution. To~begin, choose the gauge $\Psi(t) = 0$ and consider the general solution (\ref{LagEqSolutions}). This solution describes the evolution of the system from initial data 
\bse\label{initialdata}\begin{align}
    q_1(0) = -B \ ,&  \quad \dot q_1(0) = -A \ , \\
    q_2(0) = 0 \ ,&  \quad \dot q_2(0) = 0 \ , \\
    q_3(0) = C-A \ ,&  \quad \dot q_3(0) = B \ , \\
    q_4(0) = B \ ,&  \quad \dot q_4(0) = A \ .
\end{align}\ese
The configuration at some arbitrary final time $t = T$ is 
\bse\label{configwithnopsi}\begin{align}
    q_1(T) & =  -A\sin T - B\cos T \ ,\\
    q_2(T) & =  0 \ ,\\
    q_3(T) & =  C-A\cos T + B\sin T \ ,\\
    q_4(T) & =  A\sin T + B\cos T \ .
\end{align}\ese
This configuration  corresponds to a particular state of the physical~system. 

We can choose a different gauge   in Equation~(\ref{LagEqSolutions}). As~long as the new gauge satisfies $\Psi(0) = \dot\Psi(0) = 0$, the~solution  will describe 
 evolution from the same initial data (\ref{initialdata}). 
For example, with~
\be
    \Psi(t) = \frac{(\pi^2 - 4)\epsilon}{8}  \left[\cos(\pi t/T)  - 1\right] 
    + \frac{\pi \epsilon}{4} \left[ \pi t/T - \sin(\pi t/T)  \right]
\ee
where $\epsilon = {\rm const}$, 
the configuration at $t=T$ is
\bse\label{configwithpsione}\begin{align}
    q_1(T) & =  -A\sin T - B\cos T + \epsilon \ ,\\
    q_2(T) & =  -\epsilon/2 \ ,\\
    q_3(T) & =  C-A\cos T + B\sin T +\epsilon/2\ ,\\
    q_4(T) & =  A\sin T + B\cos T - \epsilon/2 \ .
\end{align}\ese
The configurations (\ref{configwithnopsi}) 
and (\ref{configwithpsione}) represent the same physical state of the system, since they evolve from the same initial~data. 

We can express this result more compactly as 
\bse\label{gaugetransftypeone}\begin{align}
    \delta q_1 & =  \epsilon \ ,\\
    \delta q_2 & =  -\epsilon/2 \ ,\\
    \delta q_3 & =  \epsilon/2 \ ,\\
    \delta q_4 & =  -\epsilon/2 \ .
\end{align}\ese
Here, $\delta q_i$ denotes the change in $q_i$ at the generic time $T$, due to the change in gauge function $\Psi(t)$. 

Here is another example. 
With~
\be
    \Psi(t) = \frac{\pi^2\epsilon}{4T} \left[ \cos(\pi t/T) - 1 \right]     + \frac{\pi\epsilon}{2T} \left[\pi t/T - \sin(\pi t/T) \right]
\ee
we obtain a configuration that differs from Equation~(\ref{configwithnopsi}) by 
\bse\label{configwithpsitwo}\begin{align}
    \delta q_1 & =  0  \ ,\\
    \delta q_2 & =   0\ ,\\
    \delta q_3 & =   \epsilon \ ,\\ 
    \delta q_4 & =  0 \ .
\end{align}\ese
This configuration is  also evolved from the initial data (\ref{initialdata}),  and~represents the same physical state as
the configurations (\ref{configwithnopsi}) and (\ref{configwithpsione}). 

Although the  gauge transformation (\ref{gaugewithepsilon}) contains a single arbitrary function of time, the~gauge invariance naturally splits into two types. The~first consists of variations subject to $\delta q_2 = -\delta q_3  =  \delta q_4 = -\delta q_1/2$. The~second consists of arbitrary variations in $\delta q_3$, 
with $\delta q_1 = \delta q_2 = \delta q_4 = 0$. This apparent ``doubling'' of the gauge freedom arises because 
the solution (\ref{gaugewithepsilon}c)  for $q_3(t)$ (unlike the other variables) includes the integral of $\Psi(t)$.  There is enough freedom of choice in $\Psi(t)$ to allow variations in $q_3$ that are independent of the variations among the other variables. Both types of gauge transformations leave the physical state of the system~unchanged. 

The consequences of gauge invariance are most clearly expressed in the Hamiltonian formalism.  The~extended Hamiltonian defined in Section~\ref{sec:extendedH} includes phase space generators for both types of gauge~transformations. 

\section{Primary Constraints and the Canonical~Hamiltonian}\label{sec:primaryconstraints}
We now begin construction of the Hamiltonian description of the system. The~conjugate momenta are defined as usual by $p_i = \partial L/\partial \dot q_i$. For~the Lagrangian (\ref{TheLagrangian}), we have 
\bse\label{momentumdefs}\begin{align}
    p_1 & =  0 \\
    p_2 & =  2\dot q_2 + \dot q_3 - \dot q_4 + q_1 \\
    p_3 & =  \dot q_2 + \dot q_3 + q_1 \\
    p_4 & =  \dot q_4 - \dot q_2
\end{align}\ese
Because the Lagrangian is singular, the~matrix of second 
derivatives $\partial^2L/\partial \dot q_i \partial \dot q_j$ is not invertible and we cannot solve  Equations~(\ref{momentumdefs}) for the velocities as functions of the coordinates and momenta. The~definitions (\ref{momentumdefs}) yield two {\em primary constraints},
\bse\label{primaryconstraints}\begin{align}
    \phi_1 & \equiv  p_1  = 0 \ , \\
    \phi_2 & \equiv  p_2 - p_3 + p_4 = 0 \ ,
\end{align}\ese
that restrict the phase space variables $p_i$, $q_i$. We will denote these constraints collectively by $\phi_a$, where $a = 1,2$. 
Note that in this simple example, the~primary constraints are independent of the $q$'s. 

There is freedom in choosing how the constraints are written. For~example, we could replace the $\phi$'s above with 
\bse\label{primeconst2}\begin{align}
\tilde\phi_1 & \equiv  p_1 \ ,\\
\tilde\phi_2 & \equiv  -p_1 + p_2 - p_3 + p_4 \ .
\end{align}\ese
In fact, any choice for the constraints is allowed, as~long as they follow from the definitions 
$p_i = \partial L/\partial \dot q_i$ and 
satisfy the {\em regularity conditions}. These  conditions state that, roughly speaking, the~constraints should have nonzero gradients on the constraint surface. More precisely, the~Jacobian matrix formed from the derivatives of the constraints with respect to the $p$'s and $q$'s should have maximal rank on the constraint subspace~\cite{DiracLectures,HenneauxTeitelboim}. For~both choices, (\ref{primaryconstraints}) and (\ref{primeconst2}), the~rank of the Jacobian matrix is $2$. On~the other hand, the~set
\bse\begin{align}
    \bar\phi_1 & \equiv  p_1^2 \ ,\\
    \bar\phi_2 & \equiv  p_2 - p_3 + p_4  \ ,
\end{align}\ese
is not permissible because the gradient of $\bar\phi_1$ vanishes on the constraint surface where \mbox{$p_1 = 0$}. 
Correspondingly, the~rank of the Jacobian matrix is less than $2$ on the constraint~surface. 

The next step in constructing the Hamiltonian formalism is to compute the {\em canonical Hamiltonian}. The~canonical Hamiltonian $H_C$ is defined from the usual prescription by writing $p_i \dot q_i - L(q,\dot q)$ in terms of $p$'s and $q$'s. Although~we cannot solve for all of the $\dot q$'s in terms of $p$'s and $q$'s,
it can be shown that the combination 
$p_i \dot q_i - L(q,\dot q)$ depends only on the phase space variables~\cite{DiracLectures,HenneauxTeitelboim}.  
For~our example problem,
  the canonical Hamiltonian is 
\be
    H_C = \frac{1}{2} \Bigl[ p_3^2 + p_4^2 - 2p_3 q_1 - (q_1 + 2q_2)(q_1+ 2q_4) \Bigr] \ .
\ee
Note that $H_C$ is ambiguous. For~example, we could use the primary constraint (\ref{primaryconstraints}b) to replace the term $-2p_3 q_1$ with $-2 (p_2 + p_4)q_1$.

\section{Primary Hamiltonian and the Primary~Action}\label{sec:primaryhamiltonian}
The {\em primary Hamiltonian} $H_P$ is obtained from the canonical Hamiltonian $H_C$ by adding the primary constraints with Lagrange multipliers,
\be
    H_P = H_C    +  \lambda_a \phi_a  \ ,
\ee
where a sum over the repeated index $a$ is implied. 
The {\em primary action} is built from the primary Hamiltonian in the usual way: $S_P[q,p,\lambda] = \int_{0}^{T} dt \{ p_i \dot q_i - H_P\}$.  Explicitly, we have
\begin{align}\label{primaryphasespaceaction}
    S_P[q,p,\lambda] 
     =  \int_{0}^{T} dt \biggl\{ &  p_i \dot q_i   
    - \frac{1}{2} \Bigl[ p_3^2 + p_4^2 - 2p_3 q_1 - (q_1 + 2q_2)(q_1 + 2q_4) \Bigr] \nono\\ 
    &  - \lambda_1 p_1 - \lambda_2(p_2 - p_3 + p_4) 
     \biggr\} \ .
\end{align}
The primary action is a functional of the complete 
set of phase space coordinates, $q_i$, $p_i$, as~well as the 
Lagrange multipliers $\lambda_1$ and $\lambda_2$. 

The equations of motion  are obtained by extremizing the primary action $S_P$. Extremization with respect to the momenta $p_i$ gives  
\bse\label{PSeoms}\begin{align}
    \dot q_1 & =  \lambda_1  \ ,\\
    \dot q_2 & =  \lambda_2 \ ,\\
    \dot q_3 & =  p_3 - q_1 - \lambda_2 \ ,\\
    \dot q_4 & =  p_4 + \lambda_2  \ ,
\end{align}
while extremization with respect to the coordinates $q_i$ yields
\begin{align}
    \dot p_1 & =  p_3  + q_1 + q_2 + q_4 \ ,\\
    \dot p_2 & =  q_1 + 2q_4 \ , \\
    \dot p_3 & =  0 \ ,\\
    \dot p_4 & =  q_1 + 2q_2 \ .
\end{align}
Extremizing the action $S_P$ with respect to the Lagrange multipliers gives the constraints, 
\begin{align}
    \phi_1 & \equiv  p_1 = 0 \ \ , \\
    \phi_2 & \equiv  p_2 - p_3 + p_4 = 0 \ .
\end{align}\ese
These equations of motion (\ref{PSeoms}) are equivalent to Lagrange's Equations~(\ref{LagrangesEqns}). To~show this, we first solve Equations~(\ref{PSeoms}c,d,i,j) for the momenta to obtain
\bse\begin{align}
    p_1 & =  0 \ ,\\
    p_2 & =  \dot q_3 - \dot q_4 + q_1 + 2\lambda_2 \ ,\\
    p_3 & =  \dot q_3 + q_1 + \lambda_2 \ ,\\
    p_4 & =  \dot q_4 - \lambda_2 \ .
\end{align}\ese
Using these results along with 
Equations~(\ref{PSeoms}a,b) for the Lagrange multipliers, we find that Equations~(\ref{PSeoms}e,f,g,h) agree precisely with Lagrange's
Equations~(\ref{LagrangesEqns}). 

\section{Hamilton's Equations and the Initial Value~Problem}\label{sec:hamiltonsequations}
At this point one  might ask whether the task of expressing the singular system (\ref{TheLagrangian}) in 
Hamiltonian form is complete. After~all, the~primary action (\ref{primaryphasespaceaction})  provides the correct equations of motion for the phase space variables $q_i$ and $p_i$. 
In fact, we can obtain the time evolution of any phase space function $F$ from $\dot F = [F,H_P]$
where $H_P$ is the primary Hamiltonian and $[\,\cdot\,  ,\,\cdot\,]$ denotes Poisson brackets. Hamilton's equations for the coordinates and momenta, $\dot q_i = [q_i,H_P]$ and $\dot p_i = [p_i,H_P]$, coincide with  Equations~(\ref{PSeoms}a--h).  

Our task of expressing the singular system in Hamiltonian form is not yet complete  because we still need to interpret Hamilton's Equations~(\ref{PSeoms}a--h) as an initial value problem. 
That is, Hamilton's equations should determine the future history of the system solely from initial data. In~contrast, the~primary action (\ref{primaryphasespaceaction})  defines a boundary value problem in which the configuration variables, the~$q$'s, 
are specified at initial and final~times. 

The key difference between the equations of motion $\delta S_P = 0$ and Hamilton's equations $\dot F = [F,H_P]$ is that the former include the primary constraints, Equations~(\ref{PSeoms}i,j), 
whereas the latter do not. 
Thus, the~phase space trajectories that extremize the action $S_P$ must lie entirely in the primary constraint surface. (The primary constraint ``surface'' is the subspace of phase space that satisfies the primary constraints.) In contrast, the~trajectories obtained from Hamiltonian evolution $\dot F = [F,H_P]$ are defined throughout the entire phase space. 
Note that we cannot simply append the primary constraint equations to Hamilton's equations, because~in that case the complete system would not be in 
Hamiltonian~form. 

Of course,
 the physically allowed phase space trajectories must satisfy the primary constraints. With~an initial value interpretation of Hamilton's equations, we can try to enforce the primary constraints with appropriate choices of initial data and Lagrange multipliers. In~particular, we can choose initial data that lie on the primary constraint surface 
$\phi_1(0) \equiv  p_1(0) = 0$ and 
$\phi_2(0)  \equiv  p_2(0) - p_3(0) + p_4(0) = 0$. 
However, 
this is not enough, because~the primary constraints are not necessarily satisfied at later times as the system evolves into the~future. 

We can describe the situation as follows. The~trajectories that extremize the action $S_P$, the~{\em physical trajectories}, do not necessarily fill the entire primary constraint surface. Instead they might span only a subspace of the primary constraint surface. If~the initial data lie in the primary constraint surface but {\em outside} the subspace of physical trajectories, then the primary constraints will not be preserved as the data are~evolved. 

How should the initial data and Lagrange multipliers be restricted such that the primary constraints hold throughout the evolution? 
The primary constraints will hold for all time if they hold initially and their time derivatives (to all orders) also vanish initially. 
In~the general case,
 this leads to a hierarchy of restrictions on the initial data in the form of secondary, tertiary, etc.,
  constraints.\footnote{Some authors use the term ``secondary constraints'' to refer to all higher--order constraints---that is, all constraints beyond the primary level.} It can also lead to restrictions on the Lagrange~multipliers.

The higher order (secondary, tertiary, etc.) constraints  and restrictions on the Lagrange multipliers are not new---imposing them does not change the content or predictions of the physical theory. 
This is because the higher-order constraints and restrictions on the Lagrange multipliers are direct consequences of the equations of motion (\ref{PSeoms}) that follow from the primary action (\ref{primaryphasespaceaction}). They are simply ``hidden'' in those equations. The~process of identifying the higher-order constraints and restrictions on Lagrange multipliers reveals these hidden~conditions. 

\section{Consistency Conditions, Secondary Constraints and Restrictions on the Lagrange~Multipliers}\label{sec:consistencyconditions}
We can ensure that the primary constraints
hold for all time by applying Dirac's {\em consistency 
conditions} \cite{DiracLectures}. Begin by computing the time derivatives of the primary constraints with the primary Hamiltonian, $\dot\phi_a = [\phi_a,H_P]$.  
Now set these equal to zero:
\be\label{ConsistencyCondition}
    [\phi_a,H_P] = 0 \ .
\ee
For each value of the index $a$, there are three possibilities.\footnote{This assumes Lagrange's equations are self--consistent. Otherwise, the~consistency conditions could lead to a contradiction such as $1 = 0$.} First, $[\phi_a,H_P]$ might vanish on the constraint surface $\phi_a=0$, so that the consistency condition (\ref{ConsistencyCondition}) reduces to the identity $0 = 0$. Second, $[\phi_a,H_P]$ could be a 
(non-constant) phase space function that is independent of the Lagrange multipliers. 
In~this case,
 Equation~(\ref{ConsistencyCondition}) is a {\em secondary constraint}. 
 Finally,  $[\phi_a,H_P]$ might depend on the Lagrange multipliers. 
 Then,
  Equation~(\ref{ConsistencyCondition}) fixes one of the Lagrange multipliers in terms of the phase space variables and the other Lagrange~multipliers. 

The secondary constraints that arise from this process must themselves satisfy the
consistency conditions. This can lead to {\em tertiary constraints} and more restrictions on the Lagrange multipliers.
 In~turn,
  the tertiary constraints can lead to quaternary constraints, and~so forth. We must continue to apply the consistency conditions until the process naturally~stops. 

For our example, the~primary constraints  are
\bse\begin{align}
    \phi_1 & \equiv  p_1 = 0\ ,\\
    \phi_2 & \equiv  p_2 - p_3 + p_4 = 0 \ ,
\end{align}\ese
and their time derivatives are 
\bse\label{firstderivativephi}\begin{align}
    \dot\phi_1 & =  [\phi_1,H_P] = p_3 +q_1 +q_2 + q_4  \ , \\
    \dot\phi_2 & =  [ \phi_2,H_P] = 2(q_1 + q_2 + q_4) \ .
\end{align}\ese
Thus, we find the secondary constraints 
\bse\begin{align}
    \psi_1 & \equiv   p_3 +q_1 +q_2 + q_4 = 0 \ ,\\
    \psi_2 & \equiv   2(q_1 + q_2 + q_4) = 0 \ .
\end{align}\ese
These will be denoted collectively by $\psi_a$. 

Applying the consistency conditions to the secondary constraints gives
\bse\label{dotpsieqns}\begin{align}
    \dot\psi_1 & =  [\psi_1,H_P] = p_4 + \lambda_1 + 2\lambda_2 = 0 \ ,\\
    \dot\psi_2 & =  [\psi_2,H_P] = 2(p_4 + \lambda_1 + 2\lambda_2) = 0 \ .
\end{align}\ese
These equations restrict the Lagrange multipliers to satisfy 
\be\label{LagMultRestriction}
    p_4 + \lambda_1 + 2\lambda_2 = 0 \ .
\ee
The process has now terminated. 
In~this example,
 there are no tertiary or higher-order~constraints.

Recall from the previous section that our goal was to restrict the initial data and Lagrange multipliers such that the primary constraints 
vanish for all times under the Hamiltonian evolution defined by $H_P$. 
We achieve this by imposing the primary constraints at the initial time, 
\bse\label{phiinitialvalues}\begin{align}
    \phi_1(0) & \equiv   p_1(0) = 0 \ ,\\
    \phi_2(0) & \equiv  p_2(0) - p_3(0) + p_4(0) = 0 \ ,
\end{align}\ese
the secondary constraints at the initial time, 
\bse\label{psiinitialvalues}\begin{align}
    \psi_1(0) & \equiv  p_3(0) +q_1(0) +q_2(0) + q_4(0) = 0 \ ,\qquad\\
    \psi_2(0) & \equiv  2[q_1(0) + q_2(0) + q_4(0) ] = 0 \ ,
\end{align}\ese
and restricting the Lagrange multipliers to satisfy Equation~(\ref{LagMultRestriction}) for {\em all} time $t$. 

Let us review the reasoning. 
From~ Equations~(\ref{dotpsieqns}), the~restriction (\ref{LagMultRestriction}) on the Lagrange multipliers tells us that $\dot\psi_a = 0$ for all time. By~Equations~(\ref{psiinitialvalues}),  $\psi_a$ vanishes initially, so we see that $\psi_a$ must vanish for all time. 
Now we use  Equations~(\ref{firstderivativephi}) to conclude that $\dot\phi_a$ must vanish for all time. Since $\phi_a$ vanishes initially, by~Equations~(\ref{phiinitialvalues}), it follows that the primary constraints $\phi_a = 0$ must hold for all time $t$. 

\section{Weak Equality and Lagrange Multiplier~Analysis}\label{sec:weakequality}
It will be useful to follow the general Dirac--Bergmann algorithm closely and carry out a formal analysis of the 
restriction (\ref{LagMultRestriction}) on the Lagrange multipliers~\cite{DiracLectures,HenneauxTeitelboim}. We begin with the concept of {\em weak equality}. 

Let ${\cal C}_A$ denote the complete set of (primary, secondary, tertiary, etc.) constraints. For~our example, 
\be
    {\cal C}_A  \equiv  \begin{pmatrix}  \phi_1 \\ \phi_2 \\\psi_1\\\psi_2 \end{pmatrix} 
    = \begin{pmatrix} p_1 \\ p_2 - p_3 + p_4 \\ p_3 + q_1 + q_2 + q_4\\ 2(q_1 + q_2 + q_4)  \end{pmatrix}
\ee
where the index $A$ runs from $1$ to $4$. 

Two phase space functions $F$ and $G$ are {\em weakly equal} if they are equal when the (primary, secondary, tertiary, etc.) constraints hold. 
In other words, $F$ and $G$ are  weakly equal if they coincide on the constraint surface, the~subspace of phase space defined by ${\cal C}_A = 0$.  Weak equality is written as $F\approx G$. 

Functions $F$ and $G$ are  {\em strongly equal} if they agree throughout phase space. Strong equality is written as $F = G$. 

Now 
 we turn to the formal analysis of the restriction (\ref{LagMultRestriction}) on the Lagrange multipliers. This restriction can be expressed as the weak equality $\dot{\cal C}_A \approx 0$. From~Equations~(\ref{firstderivativephi}) and (\ref{dotpsieqns}), we have 
\be
    \dot{\cal C}_A = [{\cal C}, H_P] = \begin{pmatrix} p_3 + q_1 + q_2 +q_4\\ 2(q_1 + q_2 + q_4) \\ p_4  + \lambda_1 + 2\lambda_2 \\ 2(p_4 + \lambda_2 + 2\lambda_2) \end{pmatrix}  \approx 
    \begin{pmatrix} 0 \\ 0 \\ 0 \\ 0 \end{pmatrix}
\ee
which simplifies to 
\be
    \begin{pmatrix} 0 \\ 0 \\ p_4 \\ 2p_4 \end{pmatrix} + 
    \begin{pmatrix} 0 & 0 \\ 0 & 0 \\ 1 & 2 \\ 2 & 4 \end{pmatrix} \begin{pmatrix} \lambda_1 \\ \lambda_2 \end{pmatrix} \approx 
    \begin{pmatrix} 0 \\ 0 \\ 0 \\ 0 \end{pmatrix} \ .
\ee
This is a system of inhomogeneous linear equations for the Lagrange multipliers. 
A particular solution is 
\be
    \left.\begin{pmatrix} \lambda_1 \\ \lambda_2 \end{pmatrix}\right|_{\rm particular} 
    = \begin{pmatrix} 0 \\ -p_4/2 \end{pmatrix} \ ,
\ee
and the homogeneous solutions are 
\be
    \left.\begin{pmatrix} \lambda_1 \\ \lambda_2 \end{pmatrix}\right|_{\rm homogeneous} 
    = \begin{pmatrix} 1 \\ -1/2 \end{pmatrix}\lambda \ ,
\ee
where $\lambda$ is arbitrary. 
The general solution is the sum of particular and homogeneous solutions: 
\be
    \left.\begin{pmatrix} \lambda_1 \\ \lambda_2 \end{pmatrix}\right|_{\rm general} =\begin{pmatrix} \lambda \\ -(\lambda + p_4)/2 \end{pmatrix}  \ .
\ee
Thus, the~restriction (\ref{LagMultRestriction}) on the Lagrange multipliers yields $\lambda_1 = \lambda$ and $\lambda_2 = -(\lambda + p_4)/2$, where $\lambda$ is an arbitrary function of~time. 

\section{Total~Hamiltonian}\label{sec:totalhamiltonian}
The {\em total Hamiltonian} $H_T$ is obtained from the primary Hamiltonian $H_P$ by inserting 
the general solution for the Lagrange multipliers: 
\begin{align}\label{totalHam}
    H_T & =   H_P\Bigr|_{\lambda_1 = \lambda,\, \lambda_2 =  -(\lambda + p_4)/2} 
    = H_C + \lambda  \phi_1 - (\lambda + p_4)\phi_2/2  \nono\\
    & =  \frac{1}{2} \left[ p_3^2 - p_2p_4 + p_3p_4 - 2p_3 q_1 
    - (q_1 + 2q_2) (q_1 + 2q_4) \right] 
    \nono\\
   & \quad + \lambda(p_1 - p_2/2 + p_3/2 - p_4/2) \ .
\end{align}
Physical phase space trajectories are defined by the  total Hamiltonian as the weak equality $\dot F \approx [F,H_T]$, with~initial data that satisfy 
the complete set of constraints, ${\cal C}_A = 0$. 

Hamilton's equations for the total Hamiltonian $H_T$ are
\bse\label{HTeqns}\begin{align}
    \dot q_1 & \approx  \lambda \ , \\
    \dot q_2 & \approx  -p_4/2 - \lambda /2 \ , \\
    \dot q_3 & \approx  p_3 + p_4/2 - q_1 + \lambda/2 \ , \\
    \dot q_4 & \approx   p_3/2 - p_2/2 -\lambda/2 \ , 
\end{align}
and
\begin{align}
    \dot p_1 & \approx  p_3 + q_1 + q_2 + q_4 \ , \\
    \dot p_2 & \approx  q_1 + 2q_4 \ , \\
    \dot p_3 & \approx  0 \ , \\
    \dot p_4 & \approx  q_1 + 2q_2 \ .
\end{align} \ese
Since these are weak equalities, we can 
use the constraints  to simplify the results. Observe that the constraints ${\cal C}_A = 0$ imply $p_1=p_3=0$, $p_2+p_4=0$ and 
$q_1+q_2+q_4=0$. 
Therefore,
 we can 
set $p_1$ and $p_3$ to zero, replace $p_4$ with $-p_2$, and~replace $q_4$ with $-q_1 - q_2$. 
Then Equations~(\ref{HTeqns}d,e,g,h) are either redundant or vacuous, and~the remaining equations 
are  
\bse\label{HTeqnsreduced}\begin{align}
    \dot q_1 & \approx  \lambda \ , \\
    \dot q_2 & \approx  p_2/2 - \lambda /2 \ , \\
    \dot q_3 & \approx  - p_2/2 - q_1 + \lambda/2 \ , \\
    \dot p_2 & \approx  -q_1 - 2q_2 \ . 
\end{align}\ese
These equations, along with the constraints ${\cal C}_A= 0$, give a complete description of the physical~system.  

Let us check the results. 
The~Lagrange multiplier $\lambda$ can be eliminated from Equations~(\ref{HTeqnsreduced}a,b) to give 
$\dot q_1 + 2\dot q_2 \approx p_2$.
Now differentiate this equation and eliminate $\dot p_2$ with Equation~(\ref{HTeqnsreduced}d) to obtain
$\ddot q_1 + 2\ddot q_2 \approx -q_1 - 2q_2$. The~
constraints allow us to set $q_1 \approx -q_2 - q_4$, which gives 
\be
    \ddot q_4 - \ddot q_2 \approx q_2 - q_4 \ .
\ee
This is Equation~(\ref{Leqn}c), which follows directly from Lagrange's equations. 
The result (\ref{Leqn}a) from Lagrange's equations is simply the secondary constraint ${\cal C}_4 \equiv 2(q_1 + q_2 + q_4) = 0$. Finally, the~result (\ref{Leqn}b)  is obtained by summing Equations~(\ref{HTeqnsreduced}b) and (\ref{HTeqnsreduced}c).  

Recall that Equations~(\ref{Leqn}a--c) are equivalent to Lagrange's equations. Thus, we have verified that Hamilton's Equations~(\ref{HTeqns}), along with the primary and secondary constraints ${\cal C}_A = 0$, are equivalent to Lagrange's~equations. 

\section{First and Second Class~Constraints}\label{sec:firstandsecondclass}
A first class function $F$ is a phase space function that has weakly vanishing Poisson brackets with all primary and secondary constraints:
\be
    [F,{\cal C}_A] \approx 0 \Longleftrightarrow F {\rm \ is\ first\ class.}
\ee
It can be shown that the Poisson bracket of any two first class functions is itself a first class function~\cite{DiracLectures,HenneauxTeitelboim}. 

The constraints themselves can be first class; constraints that are not first class are called second class. 
The constraints are separated into first and second class by examining the matrix of Poisson brackets: 
\be
    [{\cal C}_A,{\cal C}_B] = \begin{pmatrix} 
    0 & 0 & -1 & -2 \\
    0 & 0 & -2 & -4 \\
    1 & 2 & 0 & 0 \\
    2 & 4 & 0 & 0 \end{pmatrix} \ .
\ee
The rank of this $4\times 4$ matrix is $2$, and~its nullity is $4-2 = 2$. It follows that there are $2$ independent  eigenvectors with eigenvalues equal to zero; 
for example,
 $u^A = (1,-1/2,0,0)$ and  $v^A= (0,0,1,-1/2)$. 
 Then
  there are two independent combinations of constraints that are first class, namely $u^A{\cal C}_A$ and $v^A{\cal C}_A$. (A sum over the repeated index $A$ is implied.) 
The first class constraints are
\bse\begin{align}
    {\cal C}_1^{(fc)} & \equiv  \phi_1 - \phi_2/2 = p_1 - p_2/2 + p_3/2 -p_4/2  \ ,\qquad\\
    {\cal C}_2^{(fc)} & \equiv  \psi_1 - \psi_2/2 = p_3 \ .
\end{align}\ese
One can check that the first class conditions $[{\cal C}_1^{(fc)},C_B] =0$ and 
$[{\cal C}_2^{(fc)},C_B] =0$  hold. The~most general first class constraint is a linear combination of ${\cal C}_1^{(fc)}$ and ${\cal C}_2^{(fc)}$. 

There are two remaining linear combinations of constraints, which we take to be 
\bse\begin{align}
    {\cal C}_1^{(sc)} & \equiv  (\phi_1 + \phi_2)/3 = (p_1 + p_2 - p_3 + p_4)/3 \ ,\qquad \\
    {\cal C}_2^{(sc)} & \equiv  (\psi_1 + \psi_2)/3 = p_3/3 + q_1 + q_2 + q_4 \ .
\end{align}\ese
These are the second class constraints. They 
have nonvanishing Poisson brackets with each other, 
\be
    [{\cal C}_2^{(sc)}, {\cal C}_1^{(sc)} ] = 1 \ .
\ee
The most general second class constraint is a linear combination of ${\cal C}_1^{(sc)}$, ${\cal C}_2^{(sc)}$, 
${\cal C}_1^{(fc)}$ and ${\cal C}_2^{(fc)}$, with~nonzero coefficients on one or both of ${\cal C}_1^{(sc)}$ and ${\cal C}_2^{(sc)}$. 

The splitting of constraints into first and second class 
is independent of the splitting into primary and secondary. 
In~this example,
 the first class constraints are mixtures of  primary and secondary constraints. Likewise, the~second class constraints are mixtures of primary and secondary~constraints.

\section{First Class Hamiltonian, Gauge Generators and the Dirac~Conjecture}\label{sec:firstclasshamiltonian}
The total Hamiltonian (\ref{totalHam}) includes the product of an arbitrary Lagrange multiplier $\lambda$ with the first class constraint ${\cal C}_1^{(fc)} \equiv \phi_1 - \phi_2/2$. We refer to ${\cal C}_1^{(fc)}$ as a {\em primary first class constraint}, since it is constructed entirely from primary~constraints.  

If we remove the primary first class constraint from the total Hamiltonian, what remains is the {\em first class Hamiltonian} $H_{fc}$. That is, the~total Hamiltonian can be written as 
\be\label{HTsplitting}
    H_T = H_{fc} + \lambda {\cal C}_1^{(fc)} \ ,
\ee
where 
\be\label{FCHamiltonian}
    H_{fc} = \frac{1}{2} \Bigl[  p_3^2 - p_2p_4 + p_3p_4 - 2p_3 q_1 - (q_1 + 2q_2) (q_1 + 2q_4) \Bigr]  
\ee
is the first class Hamiltonian.
A common notation for $H_{fc}$, the~notation used by Dirac~\cite{DiracLectures}, is $H'$. 

We can check directly that the first class Hamiltonian (\ref{FCHamiltonian}) is a first class function.
 However, this is not necessary, because~we know that the constraints 
are preserved under the time evolution defined by $H_T$. That is, $\dot{\cal C}_A = [{\cal C}_A,H_T] \approx 0$. Thus, the~total Hamiltonian must be first class, 
$[H_T,{\cal C}_A] \approx 0$. 
Of~course,
 the primary first class constraint ${\cal C}_1^{(fc)}$ is first class. It then follows from the definition (\ref{HTsplitting}) that $H_{fc}$ must also be a first class~function. 

The splitting (\ref{HTsplitting}) of the total Hamiltonian into the first class Hamiltonian and the primary first class constraint is not special to our example. This splitting will occur for any constrained Hamiltonian system~\cite{DiracLectures,HenneauxTeitelboim}. In~general, $H_T$ will include the products of every primary first class constraint with an arbitrary~multiplier.  

Primary first class constraints generate gauge transformations. Consider the change in a phase space function $F$ generated by the primary first class constraint ${\cal C}_1^{(fc)}$,
\be\label{Gaugewithfirstclassprimary}
    \delta F = \delta\epsilon [F,{\cal C}_1^{(fc)}] \ .
\ee
This transformation does not change the physical state of the system. We can see this by considering $F$ to be evaluated as a function of the $q$'s and $p$'s at some particular time $t$. At~an infinitesimally later time $t +\delta t$, this function becomes $F(t + \delta t) = F(t) + [F,H_T]\delta t$. In~terms of the first class Hamiltonian, we have
\be\label{Fequation}
    F(t+\delta t) = F(t) + \left\{ [F,H_{fc}] + \lambda [F,C_1^{(fc)}] \right\} \delta t \ .
\ee 
The Lagrange multiplier is arbitrary, so we can make a different choice during the time interval from $t$ to $t + \delta t$, say, $\tilde\lambda$. 
Then, 
the function $F$ at time $t+\delta t$ will be  
\be\label{Fbarequation}
    \tilde F(t+\delta t) = F(t) + \left\{ [F,H_{fc}] + \tilde \lambda [F,C_1^{(fc)}] \right\} \delta t \ .
\ee
The physical state of the system at $t + \delta t$  should not depend on our choice of Lagrange multiplier, so $F(t + \delta t)$ and $\tilde F(t + \delta t)$ must represent the same physical state.  The~result (\ref{Gaugewithfirstclassprimary}) is obtained by subtracting Equation~(\ref{Fequation}) from Equation~(\ref{Fbarequation}) and defining 
$\delta F \equiv \tilde F - F$ and $\delta\epsilon \equiv (\tilde\lambda - \lambda)\delta t$. 

For the phase space coordinates, the~gauge transformation generated by the primary first class constraint ${\cal C}_1^{(fc)}$ is 
\bse\label{infinitesimalgauge}\begin{align}
    \delta q_1 & =   \delta\epsilon  \ ,\\
    \delta q_2 & =  -\delta\epsilon/2 \ ,\\
    \delta q_3 & =  \delta\epsilon/2 \ ,\\
    \delta q_4 & =  -\delta\epsilon/2 \ .
\end{align}\ese
The transformations of the $p$'s all vanish. This result agrees with the gauge transformation from Equation~(\ref{gaugetransftypeone}), with~the change of notation $\epsilon \leftrightarrow \delta\epsilon$. Here, we denote the gauge parameter by $\delta\epsilon$ because the transformation is infinitesimal; in Equation~(\ref{gaugetransftypeone}),
 we used $\epsilon$ because the transformation was finite. It is clear that the infinitesimal transformation (\ref{infinitesimalgauge}) can be iterated to obtain the finite transformation (\ref{gaugetransftypeone}). 

In general, a~gauge transformation is defined as a transformation 
$\delta F = \delta\epsilon[F,{\mathbb G}]$ that does not alter the physical state of the system. The~function ${\mathbb G}$ is the gauge generator. We have seen that the primary first class constraints generate gauge transformations. 
However, not all gauge transformations are generated by primary first class constraints. 
In~fact, it can be shown~\cite{DiracLectures,HenneauxTeitelboim} that the Poisson bracket between any primary first class constraint and the first class Hamiltonian is itself a first class constraint that generates a gauge transformation.
\footnote{For systems with more than one primary first class constraint, the~Poisson 
brackets of any two primary first class constraints is also a first class constraint that generates a gauge transformation~\cite{DiracLectures,HenneauxTeitelboim}.}

For our example problem, the~Poisson bracket of the primary first class constraint ${\cal C}_1^{(fc)}$ and the first class Hamiltonian $H_{fc}$ is
\be
    [{\cal C}^{(fc)}_1,H_{fc}] = p_3 \ .
\ee
This is the secondary first class constraint, ${\cal C}_2^{(fc)} \equiv \psi_1 - \psi_2/2 = p_3$. 
Thus, we see that in this example both the primary and secondary first class constraints are generators of gauge  transformations. 
Explicitly, the~transformation $\delta F = \delta\epsilon[F,{\cal C}_2^{(fc)}]$ is 
\bse\label{infinitesimalgaugetypetwo}\begin{align}
     \delta q_1 & =   0  \ ,\\
    \delta q_2 & =  0 \ ,\\
    \delta q_3 & =  \delta\epsilon \ ,\\
    \delta q_4 & =  0 \ ,
\end{align}\ese
with the transformations of the $p$'s all vanishing.  We can iterate this infinitesimal gauge transformation 
(\ref{infinitesimalgaugetypetwo}) to obtain the finite transformation (\ref{configwithpsitwo}).

The ``doubling'' of the gauge freedom identified in Section~\ref{sec:gaugeinvariance} appears quite naturally in the Hamiltonian formalism. The~two types of gauge transformation are generated by the two first class constraints, ${\cal C}_1^{(fc)}$ and ${\cal C}_2^{(fc)}$. 

The Dirac conjecture~\cite{DiracLectures} says that {\em all} first class constraints (whether they are primary, secondary, etc., or~a combination of primary, secondary, etc.)  generate gauge transformations. This conjecture does not hold as a general theorem---there are known examples in which the transformation generated by a secondary first class constraint does not coincide with any invariance of the original Lagrangian system.\footnote{Counterexamples to the Dirac conjecture are discussed in Refs.~\cite{Cawley,HenneauxTeitelboim,Li_1993,WangLiWang,Frenkel:1980nt,Sugano:1983kc,Wu:1994jd,Miskovic:2003ex} and elsewhere. Proofs of the conjecture have been constructed by adopting various simplifying assumptions~\cite{Costa:1985tt,Cabo:1990jk,HenneauxTeitelboim}. 
The status of the conjecture is a subtle issue; see for example Refs.~\cite{Rothe:2004jc,RotheRothe,Pons}.} Nevertheless, the~Dirac conjecture is usually taken as an assumption. It appears that in practice, for~systems of physical interest, all first class constraints generate gauge~transformations.

\section{Extended~Hamiltonian}\label{sec:extendedH}
The Dirac conjecture tells us that all  
first class constraints generate gauge transformations and should be treated on an equal footing.  
The {\em extended Hamiltonian} $H_E$ is defined by adding all  first class constraints ${\cal C}_a^{(fc)}$ with Lagrange multipliers $\gamma_a$ to the first class Hamiltonian: 
\be
    H_E = H_{fc} + \gamma_a {\cal C}_a^{(fc)} \ .
\ee
(A sum over the index $a$ is implied.) 

For our example, the~extended Hamiltonian is 
\begin{align}\label{TheExtendedH}
   H_E = &  \frac{1}{2} \Bigl[ p_3^2 - p_2p_4 + p_3p_4 - 2p_3 q_1 
    - (q_1 + 2q_2) (q_1 + 2q_4) \Bigr]  \nono\\
    & + \gamma_1 (p_1 - p_2/2 + p_3/2 - p_4/2) + \gamma_2 p_3 \ .
\end{align}
The 
equations of motion $\dot F \approx [F,H_E]$ are
\bse\label{HEeqns}\begin{align}
    \dot q_1 & \approx  \gamma_1 \ ,\\
    \dot q_2 & \approx  -p_4/2 - \gamma_1/2 \ ,\\
    \dot q_3 & \approx  p_3 + p_4/2 - q_1 + \gamma_1/2 + \gamma_2 \ ,\\
    \dot q_4 & \approx  p_3/2 - p_2/2  - \gamma_1/2 \ ,
\end{align}
and
\begin{align}
    \dot p_1 & \approx  p_3 + q_1 + q_2 + q_4 \ ,\\
    \dot p_2 & \approx  q_1 + 2q_4 \ ,\\
    \dot p_3 & \approx  0 \ ,\\
    \dot p_4 & \approx  q_1 + 2q_2 \ .
\end{align}\ese
Let us compare these results to the equations of motion (\ref{HTeqns}) obtained from the total Hamiltonian $H_T$. There are just two differences. The~first is trivial: the Lagrange multiplier $\lambda$ in Equations~(\ref{HTeqns})
has changed names to $\gamma_1$ in Equations~(\ref{HEeqns}). The~second 
difference is significant: 
the equation for $\dot q_3$ has an extra term $\gamma_2$ on the right-hand side. This is a 
new feature of the extended Hamiltonian. It makes explicit the fact that the gauge freedom allows $q_3$ to be changed arbitrarily, and~independently, from~the other~variables. 

We can check the equations of motion for $H_E$ following the same reasoning that was applied to the equations of motion for $H_T$. 
First, recall that the (first and second class) constraints imply $p_1 = p_3 = 0$, $p_4 = -p_2$ and $q_4 = -q_1 - q_2$.
 Then  Equations~(\ref{HEeqns}e) and (\ref{HEeqns}g) are vacuous, and~Equations~(\ref{HEeqns}f) and (\ref{HEeqns}h) are redundant. It also follows that with the constraints imposed, Equation~(\ref{HEeqns}d) is a consequence of Equations~(\ref{HEeqns}a) and (\ref{HEeqns}b). The~remaining equations are 
\bse\label{HEeqns2}\begin{align}
    \dot q_1 & \approx  \gamma_1 \ ,\\
    \dot q_2 & \approx  p_2/2 - \gamma_1/2 \ ,\\
    \dot q_3 & \approx  -p_2/2 - q_1 + \gamma_1/2 + \gamma_2 \ ,\\
    \dot p_2 & \approx  -q_1 - 2q_2 \ .
\end{align}\ese
These agree with Equations~(\ref{HTeqnsreduced}), apart from the change of notation $\lambda \to \gamma_1$ and the extra term $\gamma_2$ on the right-hand side of the $\dot q_3$ equation. 

By eliminating $\gamma_1$, the~equations of motion generated by the extended Hamiltonian $H_E$ become 
\bse\label{HEreducedeqns}\begin{align}
    \dot q_1 + 2\dot q_2 & \approx  p_2 \ ,\\
    \dot q_2 + \dot q_3 & \approx  -q_1 + \gamma_2 \ ,\\
    \dot p_2 & \approx  -q_1 - 2q_2 \ .
\end{align}\ese
If we differentiate the first equation, combine with the second, and~use the constraint $q_1 + q_2 + q_4 = 0$, we obtain $(\ddot q_4 - \ddot q_2) \approx -(q_4 - q_2)$. This is the expected result (\ref{Leqn}c). In~fact, the~only difference between Hamilton's equations $\dot F \approx [F,H_E]$ and the results (\ref{Leqn}) (which  are equivalent to Lagrange's equations) 
is the extra term $\gamma_2$ in Equation~(\ref{HEreducedeqns}b) above. That term does not appear in the corresponding Lagrangian Equation~(\ref{Leqn}b). 

Note that we can use the first class constraints to simplify the extended Hamiltonian $H_E$. For~example, using $C_2^{(fc)} = p_3$, we can set $p_3 = 0$ everywhere in Equation~(\ref{TheExtendedH}), except~of course in the term $\gamma_2 p_3$. The~extended Hamiltonian becomes
\be\label{NewExtendedH}
	H_E =    \frac{1}{2} \Bigl[- p_2p_4 
    - (q_1 + 2q_2) (q_1 + 2q_4) \Bigr] 
      + \gamma_1 (p_1 - p_2/2  - p_4/2) + \gamma_2 p_3 \ .
\ee
This amounts to replacing the Lagrange multiplier $\gamma_2$ in Equation~(\ref{TheExtendedH}) by 
\be
    \gamma_2 \to \gamma_2 - p_3/2 - p_4/2 + q_1 - \gamma_1/2 \ .
\ee
This replacement does not change the physical content of the theory, since the Lagrange multiplier $\gamma_2$ is~arbitrary. 

\section{Extended~Action}\label{sec:extendedS}
The equations of motion for the extended theory can be derived from the action~\cite{HenneauxTeitelboim}
\be\label{ExtendedAction}
    S_E[q,p,\gamma,\sigma] = \int_{0}^{T} dt \Bigl\{ 
    p_i  \dot q_i 
     - H_E - \sigma_1 \, {\cal C}_1^{(sc)} - \sigma_2\, {\cal C}_2^{(sc)} \Bigr\} \ ,
\ee
which includes the second class constraints with Lagrange multipliers $\sigma_a$. Recall that the first class constraints ${\cal C}_a^{(fc)}$ are included in the extended Hamiltonian $H_E$ with multipliers $\gamma_a$, so $S_E$ includes all four~constraints. 

We can use either form of the extended Hamiltonian, Equation~(\ref{TheExtendedH}) or (\ref{NewExtendedH}), in~the extended action. 
Let us  use Equation~(\ref{TheExtendedH}). 
Then,
 the equations of motion that follow from extremizing $S_E$ with respect to the momenta $p_i$ are 
\bse\label{SEeoms}\begin{align}
     \dot q_1 & =  \gamma_1 + \sigma_1/3 \ ,\\
    \dot q_2 & =  -p_4/2 - \gamma_1/2 + \sigma_1/3 \ ,\qquad \\
    \dot q_3 & =  p_3 + p_4/2 - q_1 + \gamma_1/2 + \gamma_2 
   - \sigma_1/3 + \sigma_2/3 \ ,  \\
    \dot q_4 & =  p_3/2 - p_2/2 -\gamma_1/2 + \sigma_1/3 \ .
\end{align}
Extremizing $S_E$ with respect to the coordinates yields
\begin{align}
    \dot p_1 & =  p_3 + q_1 + q_2 + q_4 - \sigma_2 \ ,\\
    \dot p_2 & =  q_1 + 2q_4  - \sigma_2 \ ,\\
    \dot p_3 & =  0 \ ,\\
    \dot p_4 & =  q_1 + 2q_2 - \sigma_2 \ ,
\end{align}
and the constraints 
\begin{align}
    {\cal C}_1^{(fc)} & \equiv  p_1 - p_2/2 + p_3/2 - p_4/2 =0  \ , \\
     {\cal C}_2^{(fc)} & \equiv  p_3 = 0 \ ,\\
     {\cal C}_1^{(sc)} & \equiv  (p_1 + p_2 - p_3 + p_4)/3 = 0 \ ,\\
     {\cal C}_2^{(sc)} & \equiv  p_3/3 + q_1 + q_2 + q_4 = 0\ ,
\end{align}\ese
follow from extremizing  $S_E$ with respect to the Lagrange multipliers $\gamma_a$ and $\sigma_a$. 

Let us check these equations of motion. The~constraints imply 
$p_1 = p_3 = 0$, $p_4 = -p_2$ and $q_1 + q_2 + q_4 =0$.
Then 
 the equation of motion (\ref{SEeoms}e) gives $\sigma_2 \approx 0$.
The constraints also imply $\dot q_1 + \dot q_2 + \dot q_4 = 0$. The~sum of Equations~(\ref{SEeoms}a), (\ref{SEeoms}b)
and (\ref{SEeoms}d) then yields $\sigma_1 
\approx 0$.  Now, if~we set $\sigma_1$ and $\sigma_2$ to zero, 
the equations of motion (\ref{SEeoms}a--h) agree precisely with Hamilton's Equations~(\ref{HEeqns}) for the extended Hamiltonian $H_E$. 
In~Section~\ref{sec:extendedH},
 we showed that the equations generated by $H_E$ 
agree with Lagrange's equations apart from the extra term $\gamma_2$ in 
the equation for $\dot q_3$. This term extends the original Lagrangian theory by making 
explicit the fact that the gauge freedom allows for independent transformations of $q_3$. 

Finally, we note that the extended action is invariant under the transformation defined by 
\bse
\be
     \delta F  =  \epsilon_1[F,C^{(fc)}_1] 
     + \epsilon_2[F,C^{(fc)}_2] 
\ee
for the phase space variables and 
\begin{align}
    \delta \gamma_1 & =  \dot \epsilon_1 \ ,\\
    \delta\gamma_2 & =  \dot \epsilon_2 + \epsilon_1 \ ,\\
    \delta\sigma_1 & =  0 \ ,\\
    \delta\sigma_2  & =  0 \ ,
\end{align}\ese
for the Lagrange multipliers.  Here, the~gauge parameters $\epsilon_1$ and $\epsilon_2$ are functions of time. These equations express the gauge invariance at the level of the action $S_E$. 

\section{Dirac Brackets and the Partially Reduced~Hamiltonian}\label{ref:diracbrackets}
We now return to the evolution defined by the extended Hamiltonian $H_E$ of Equation~(\ref{TheExtendedH}), and~Hamilton's Equations~(\ref{HEeqns}). To~obtain a physically allowed trajectory, we must choose initial data that satisfy the four constraints ${\cal C}_a^{(fc)} = 0$ 
and ${\cal C}_a^{(sc)} = 0$. Apart from  restricting the initial data, the~second class constraints  play no role in the formalism. It would be convenient if we could restrict the variables from the outset such that the second class constraints are automatically satisfied. For~example, we could use ${\cal C}_1^{(sc)}  = 0$ and ${\cal C}_2^{(sc)} =0$ from Equations~(\ref{SEeoms}k,l) to replace  $q_1$ with $-q_2 - q_4 - p_3/3$ and replace $p_2$ with $-p_1 + p_3 - p_4$.  

We are not allowed to apply the second class constraints in this way. For~example, consider the Poisson brackets $[q_1,p_1] = 1$. If~we were to replace $q_1$ with $-q_2 - q_4 - p_3/3$, we would find a different answer: $[-q_2 - q_4 - p_3/3,p_1] = 0$. The~second class constraints cannot be imposed before Poisson brackets are~computed. 

Dirac devised a way to allow the second class constraints to be imposed 
from the outset  by modifying the Poisson  brackets~\cite{DiracLectures}. The~result is the Dirac~brackets. 

To construct Dirac brackets, we first 
compute the matrix of Poisson brackets among the second class constraints: 
\be
    M_{ab} \equiv [{\cal C}^{(sc)}_a ,{\cal C}^{(sc)}_b ] = 
    \begin{pmatrix} 0 & -1 \\ 1 & 0 \end{pmatrix} \ .
\ee
Let 
\be
    M^{ab} = \begin{pmatrix} 0 & 1 \\ -1 & 0 \end{pmatrix}
\ee
denote the inverse of $M_{ab}$. 
Then, 
the Dirac brackets $[F,G]^*$ of two phase space functions $F$ and $G$ are defined  by 
\be
    [F,G]^* \equiv [F,G] - [F,{\cal C}^{(sc)}_a ] M^{ab} 
    [{\cal C}^{(sc)}_b,G] \ .
\ee
Dirac brackets, like Poisson brackets,  are antisymmetric and satisfy the Jacobi identity~\cite{DiracLectures,HenneauxTeitelboim}.

Explicitly, the~Dirac brackets among the coordinates are
\be
    [q_i,q_j]^* = \frac{1}{9} \begin{pmatrix} 0 & 0 & 1 & 0 \\ 0 & 0 & 1 & 0 \\ -1 & -1 & 0 & -1 \\ 0 & 0 & 1 & 0 \end{pmatrix} \ ,
\ee
and the Dirac brackets between the $q$'s and $p$'s are 
\be
    [q_i,p_j]^* = \frac{1}{3} \begin{pmatrix} 2 & -1 & 0 & -1 \\ -1 & 2 & 0 & -1 \\ 1 & 1 & 3 & 1 \\ -1 & -1 & 0 & 2 \end{pmatrix}
\ee
For our example,
 the Dirac brackets among the momenta all vanish: $[p_i,p_j]^* = 0$. 

There are two key properties that make Dirac brackets relevant. First, the~Dirac brackets agree weakly with Poisson brackets if one of the two functions is first class.  Since  the extended Hamiltonian is first class, we have $[F,H_E]^* \approx [F,H_E]$ for any $F$. 
It follows that we can write the equations of motion as 
\be
    \dot F \approx [F,H_E]^* \ ,
\ee
using Dirac~brackets. 

The second key property of the Dirac brackets is that they weakly vanish if one of the functions is a second class constraint: $[F,{\cal C}_a^{(sc)}]^* \approx 0$.  This allows us to apply the second class constraints  before computing brackets. For~example, 
we can use either $q_1$ or $-q_2 - q_4 - p_3/3$ to compute Dirac brackets with $p_1$: 
\be
    [q_1,p_1]^* = [-q_2-q_4-p_3/3,p_1]^* = 2/3 \ .
\ee
With Dirac brackets, the~second class constraints can be treated as strong equations and imposed before computing the equations of~motion. 

Let us use the second class constraints (\ref{SEeoms}k,l) 
to eliminate $q_1$ and $p_2$ and write the extended Hamiltonian in terms of the smaller set of variables $q_2$, $q_3$, $q_4$, $p_1$, $p_3$ and $p_4$. Setting
\bse\label{scconstraintrelations}\begin{align}
    q_1 & =  -q_2-q_4-p_3/3  \ ,\\
    p_2 & =  - p_1 + p_3 -p_4 \ ,
\end{align}\ese
we have
\be
    H_{PR} = 7p_3^2/9 + \frac{1}{2} \left[ p_4^2 + p_1 p_4 + 2p_3(q_2 + q_4) + (q_2 - q_4)^2\right] + \gamma_1(3p_1/2) + \gamma_2 p_3 \ .\\
\ee
This is the {\em partially reduced Hamiltonian}, obtained from the 
extended Hamiltonian by applying the second class~constraints. 

Of course,
 the partially reduced Hamiltonian is not unique.  We could use the second class constraints to eliminate some other pair of variables instead of $q_1$ and $p_2$. 

The equations of motion generated by the partially reduced Hamiltonian, $\dot F \approx [F,H_{R}]^*$, are
\bse\label{HReqns}\begin{align}
    \dot q_2 & \approx  -p_1/6 - p_4/2 - \gamma_1/2 \ ,\\
    \dot q_3 & \approx  4p_3/3 +p_1/6 + p_4/2 + q_2 + q_4 
    + \gamma_1/2 + \gamma_2 \ ,\\
    \dot q_4 & \approx  p_1/3 + p_4/2  - \gamma_1/2 \ ,\\
    \dot p_1 & \approx  2p_3/3 \ ,\\
    \dot p_3 & \approx  0 \ ,\\
    \dot p_4 & \approx  -p_3/3 + q_2 - q_4 \ .
\end{align}\ese
We can also use $H_{PR}$ and the Dirac brackets to compute $\dot q_1$ and $\dot p_2$. The~results are 
equivalent to those obtained by differentiating the right-hand sides of Equations~(\ref{scconstraintrelations}) and using the equations of motion (\ref{HReqns}).  

Let us
 check the equations of motion. With~the second class constraints applied, the~
first class constraints imply $p_1 =p_3 = 0$ and $p_4 = -p_2$. Thus,  
Equations~(\ref{HReqns}d) and (\ref{HReqns}e) are vacuous and the remaining equations become  
\bse\label{HReqnssimp}\begin{align}
    \dot q_2 & \approx  p_2/2 - \gamma_1/2 \ ,\\
    \dot q_3 & \approx    -p_2/2  +q_2 + q_4 + \gamma_1/2 + \gamma_2 \ ,\\
    \dot q_4 & \approx   -p_2/2  - \gamma_1/2 \ ,\\
    \dot p_2 & \approx  -q_2 + q_4 \ ,
\end{align}\ese
Compare these to the independent Equations~(\ref{HEeqns2}) that follow from the extended Hamiltonian. Equations~(\ref{HEeqns2}b,c,d) agree with Equations~(\ref{HReqnssimp}a,b,d) once we use 
$q_1 = -q_2 - q_4$. The~final Equation~(\ref{HEeqns2}a) is obtained by differentiating $q_1= - q_2 -  q_3$ in time and using Equations~(\ref{HReqnssimp}a) and (\ref{HReqnssimp}b). 

\section{Partially Reduced~Action}\label{sec:partiallyreducedS}
The partially reduced equations of motion (\ref{HReqns}) can be obtained from the extended action $S_E$ by eliminating the superfluous variables. Note that the equations of motion obtained by varying 
$S_E$ with respect to $p_2$, $q_1$, $\sigma_1$ and $\sigma_2$  are Equations~(\ref{SEeoms}b,e,k,l), respectively. We can eliminate these variables by solving these equations and substituting the results into the action.\footnote{Any action can be reduced by using the equations of motion obtained by varying with respect to a subset of variables, solving those equations for the same subset of variables, then substituting the results into the action. In~general, it is not permissible to reduce an action by using the equations obtained by varying with respect to one subset of variables but solving those equations for a different subset of variables.}
The results are:
\bse\begin{align}
    q_1 & =  -p_3/3 - q_2 - q_4 \ ,\\
    p_2 & =  -p_1 + p_3 - p_4 \ ,\\
    \sigma_1 & =  3\dot q_2 + 3p_4/2 + 3\gamma_1/2 \ ,\\
    \sigma_2 & =  -\dot p_1 + 2p_3/3 \ .
\end{align}\ese
Inserting these into the extended action (\ref{ExtendedAction}), 
we find 
\begin{align}\label{partiallyreducedaction}
    S_{PR}[q_2,q_3,q_4,p_1,p_3,p_4,\gamma_1,\gamma_2] = \int_0^T dt \bigl\{ & -p_1\dot p_3/3 + (p_3 - p_4 - 2p_1)\dot q_2 \nono\\
    & + p_3 \dot q_3 + (p_4 - p_1)\dot q_4 - H_{PR} \bigr\} \ .
\end{align}
This is the partially reduced~action. 

The equations of motion obtained from varying $S_{PR}$ with respect to the phase space variables are 
\bse\label{EquationForSR}\begin{align}
    2\dot p_1 - \dot p_3 + \dot p_4 - p_3 - q_2 + q_4 & =  0 \ , \\
    -\dot p_3  & =  0 \ , \\
    \dot p_1 - \dot p_4 - p_3 + q_2 - q_4  & =  0 \ , \\
    -2\dot q_2 - \dot q_4 - \dot p_3/3 - p_4/2 - 3\gamma_1/2 & =  0 \ , \\
    \dot q_2 + \dot q_3 + \dot p_1/3 - 14p_3/9 - q_2 - q_4 - \gamma_2 & =  0 \ ,\qquad \\
    \dot q_2 + \dot q_4 - p_1/2 - p_4 & =  0
    \ ,
\end{align}\ese
and the equations obtained by varying with respect to the Lagrange multipliers $\gamma_1$ and $\gamma_2$ are
\bse\begin{align}
    -3p_1/2 & =  0  \ , \\
    -p_3 & =  0  \ .
\end{align}\ese
These are, of course,
 the first class constraints, reduced by using the second class constraints to eliminate $q_1$ and $p_2$. 

We can now solve  Equation~(\ref{EquationForSR}) for the time derivatives of $q_2$, $q_3$, $q_4$, $p_1$, $p_3$ and $p_4$. The~result coincides with the equations of motion (\ref{HReqns}) obtained from the partially reduced Hamiltonian $H_{PR}$ and the Dirac~brackets. 

The partially reduced action $S_{PR}$ is invariant under the transformation defined by 
\bse
\be
     \delta F  =  \epsilon_1[F,C^{(fc)}_1]^* 
     + \epsilon_2[F,C^{(fc)}_2]^* 
\ee
for the phase space variables and 
\begin{align}
    \delta \gamma_1 & =  \dot \epsilon_1 \ ,\\
    \delta\gamma_2 & =  \dot \epsilon_2 + \epsilon_1 \ ,
\end{align}\ese
for the Lagrange multipliers. These equations express the gauge invariance of the theory at the level of the action principle with the second class constraints~eliminated.  

\section{Partially Reduced~Lagrangian}\label{sec:partiallyreducedL}
It is not too difficult to find a change of variables that will bring $S_{PR}$ into ``canonical form''. For~example, let 
\bse\label{qptoQP}\begin{align}
    q_2 & =  Q_1 - P_2/9\ ,\\
    q_3 & =  Q_2 + P_2/9 \ ,\\
    q_4 & =  Q_3 - P_2/9 \ ,\\
    p_1 & =  (-P_1 + P_2 - P_3)/3 \ ,\\
    p_3 & =  P_2 \ ,\\
    p_4 & =  (-P_1 + P_2 + 2P_3)/3 \ ,
\end{align}\ese
define a new set of variables $Q_\alpha$, $P_\alpha$ for the secondary constraint surface. (The index $\alpha$ ranges over $1$, $2$ and $3$.) The partially reduced action becomes 
\be\label{SPRofQP}
    S_{PR}[Q,P,\gamma] = \int_0^T dt \left\{ P_\alpha \dot Q_\alpha - H_{PR} \right\}
\ee
with 
\begin{align}\label{HPRofPQ}
    H_{PR} = & \frac{1}{18}\left[ 2P_1^2 + 12 P_2^2 + 2P_3^2 - 4P_1 P_2 + 5(P_2 - P_1)P_3 + 18P_2(Q_1 + Q_3) + 9(Q_1 - Q_3)^2 \right] \nono\\
    &   + \gamma_1(-P_1 + P_2 - P_3)/2 + \gamma_2 P_2 \ .
\end{align}
The equations of motion $\delta S_{PR}=0$ include $\dot Q_\alpha = [Q_\alpha,H_{PR}]$ and $\dot P_\alpha = [P_\alpha,H_{PR}]$, where $[\,\cdot\,  , \,\cdot\, ]$ are the usual Poisson brackets. 
The partially reduced action (\ref{SPRofQP}) is invariant under the transformation 
\bse\begin{align}
     \delta F  & =    \epsilon_1[F,C^{(fc)}_1] + \epsilon_2[F,C^{(fc)}_2]  \ ,\\
     \delta \gamma_1 & =  \dot \epsilon_1 \ ,\\
    \delta\gamma_2 & =  \dot \epsilon_2 + \epsilon_1 \ ,
\end{align}\ese
where $C_1^{(fc)} = (-P_1 + P_2 - P_3)/2$ and 
$C_2^{(fc)} = P_2$ are the first class~constraints. 

We can use the equations of motion $\delta S_{PR}/\delta P_\alpha = 0$ to eliminate the momenta $P_\alpha$ from the partially reduced action (\ref{SPRofQP}). These equations are 
\bse\begin{align}
   \dot Q_1 & =  (4P_1 - 4P_2 - 5P_3)/18 - \gamma_1/2 \ ,\\
   \dot Q_2 & =  -(4P_1 - 24 P_2 - 5P_3)/18  + Q_1 + Q_3 + \gamma_1/2 + \gamma_2 \ ,\\
   \dot Q_3 & =  -(5P_1 - 5P_2 - 4P_3)/18 - \gamma_1/2 \ ,
\end{align}\ese
with solutions
\bse\begin{align}
    P_1 & =  (-71\dot Q_1 + 9\dot Q_2 - 100\dot Q_3)/10 - 9(Q_1 + Q_3)/10  - 9\gamma_1 - 9\gamma_2/10 \ ,\\
    P_2 & =  9(\dot Q_1 + \dot Q_2 - Q_1 - Q_3 - \gamma_2)/10 \ ,\\
    P_3 & =  -10\dot Q_1 - 8\dot Q_3 - 9\gamma_1 \ .
\end{align}\ese
Inserting these results into the partially reduced action, we find 
\begin{align}\label{SPRofQgamma}
    S_{PR}[Q,\gamma] = \int_0^T dt \biggl\{ & (-71\dot Q_1^2 + 9\dot Q_2^2 - 80\dot Q_3^2)/20 + (9\dot Q_2 - 100\dot Q_3)\dot Q_1/10 \nono \\ 
    & - 9(Q_1 + Q_3) (\dot Q_1 + \dot Q_2)/10 
    - (Q_1^2 + Q_3^2 - 38 Q_1 Q_3)/20   \nono\\ 
    & - 9(\dot Q_1 + \dot Q_3)\gamma_1 - 9(\dot Q_1 + \dot Q_2 - Q_1 - Q_3)\gamma_2/10  \nono\\
    & - 9(10\gamma_1^2 - \gamma_2^2)/20
    \biggr\} \ .
\end{align}
The {\em partially reduced Lagrangian} is the integrand of this~action. 

We can  go one step further and eliminate the 
Lagrange multipliers using the  equations of motion $\delta S/\delta\gamma_1 = 0$ and $\delta S/\delta\gamma_2 = 0$. These equations have solutions
\bse\begin{align}
    \gamma_1 & =  -\dot Q_1 - \dot Q_3 \ ,\\
    \gamma_2 & =  \dot Q_1 + \dot Q_2 - Q_1 - Q_3 \ .
\end{align}\ese
Inserting these results back into the action yields 
\be\label{PRactionofQ}
    S_{PR}[Q] = \int_0^T dt \left\{ \frac{1}{2} (\dot Q_1 - \dot Q_3)^2  - \frac{1}{2}(Q_1 - Q_3)^2 \right\} \ .
\ee
This is the action for a  harmonic oscillator in the variable $Q_1 - Q_3$. 
Note that the coordinate transformation (\ref{qptoQP}) implies $(Q_1 - Q_3) = (q_2 - q_4)$, so once again, we find that the coordinate combination $q_2 - q_4$ describes a simple harmonic oscillator. 
In addition, we observe that the action  (\ref{PRactionofQ}) leaves the variable $Q_2$ completely unspecified. This expresses the gauge freedom  
generated by the first class constraint $C_2^{(fc)} =  P_2 = p_3$.

\section{Gauge Conditions and the Fully Reduced~Hamiltonian}\label{sec:gaugeconditions}
Let us return to the theory described by the extended Hamiltonian, prior to the  elimination of the second class~constraints. 

For our example problem, phase space is eight-dimensional. 
The~physical trajectories fill the constraint ``surface'', which is the four-dimensional subspace where all first and second class constraints hold.  Each point in the constraint surface can be mapped into a physically equivalent state by the gauge generators, namely, the~first class constraints ${\cal C}_a^{(fc)}$. Since there are two independent gauge generators, each physical state of the system corresponds to a two-dimensional subspace of the constraint surface. The~constraint surface is foliated by these two-dimensional slices, referred to as gauge ``orbits''. 

We can select a single phase space point on each gauge orbit to represent the physical state. We do this by applying gauge conditions. 
In~particular,
 we will consider a {\em canonical gauge}\footnote{Canonical gauges restrict the phase space variables. Noncanonical gauges~\cite{HenneauxTeitelboim} involve the Lagrange multipliers.} which takes the form ${\cal G}_a(q,p) \approx 0$ with $a = 1,2$. A~good canonical gauge condition must not be gauge invariant, otherwise it would allow more than one point on the gauge orbit to represent the physical state of the system. To~be precise, the~matrix  of Poisson brackets of gauge conditions and gauge generators, $[{\cal G}_a, C_b^{(fc)}]$, must be nonsingular~\cite{HenneauxTeitelboim}.

As an example, let us choose 
\bse\begin{align}
    {\cal G}_1 & =  q_1 - q_2 \ ,\\
    {\cal G}_2 & =  q_3 + p_4 
\end{align}\ese
as our gauge conditions. 
This is a good gauge:
\be
    {\rm det}[{\cal G}_a, C_b^{(fc)}] = \begin{vmatrix} 3/2 & 0 \\ 1/2 & 1 \end{vmatrix} = 3/2 \ . 
\ee
The matrix $[{\cal G}_a, C_b^{(fc)}]$ is  nonsingular, as~required. 

The gauge conditions ${\cal G}_a = 0$, like the first and second class constraints, restrict the phase space variables. 
The full set of restrictions
\be
    {\cal C}^{(all)}_A = \{ {\cal G}_1, {\cal G}_2, {\cal C}^{(fc)}_1, {\cal C}^{(fc)}_2, {\cal C}^{(sc)}_1, {\cal C}^{(sc)}_2 \}
\ee
reduces the available phase space from eight dimensions to two dimensions. (Here, the~index $A$ ranges from $1$ to $6$.) Taken as a whole, the~six conditions ${\cal C}^{(all)}_A$ are second class. We see this by computing the Poisson brackets
\be
    M_{AB} \equiv [ {\cal C}^{(all)}_A, {\cal C}^{(all)}_B ] 
    = \frac{1}{6}\begin{pmatrix}
    0 & 0 & 9 & 0 & 0 & 0 \\
    0 & 0 & 3 & 6 & -2 & -4 \\
    -9 & -3 & 0 & 0 & 0 & 0 \\
    0 & -6 & 0 & 0 & 0 & 0 \\
    0 & 2 & 0 & 0 & 0 & -6 \\
    0 & 4 & 0 & 0 & 6 & 0
    \end{pmatrix} \ .
\ee
This matrix has a nonzero determinant, $\det(M) = 9/4$, which is the condition for the set of constraints and gauge conditions to be second~class. 

We can eliminate the constraints and gauge conditions by constructing Dirac brackets. The~inverse of $M_{AB}$ is
\be
    M^{AB} = \frac{1}{3} \begin{pmatrix}
    0 & 0 & -2 & 1 & 0 & 0 \\
    0 & 0 & 0 & -3 & 0 & 0 \\
    2 & 0 & 0 & 0 & 0 & 0 \\
    -1 & 3 & 0 & 0 & -2 & 1 \\
    0 & 0 & 0 & 2 & 0 & 3 \\
    0 & 0 & 0 & -1 & -3 & 0 
    \end{pmatrix} \ ,
\ee
and the Dirac brackets are defined by 
\be
    [F,G]^* = [F,G] - [F,{\cal C}^{(all)}_A] M^{AB} [{\cal C}^{(all)}_B,G]  \ .
\ee
The Dirac brackets among the phase space variables are 
\bse\begin{align}
    \left[q_1,q_3\right]^*  = [q_1,p_2]^* = -[q_1,p_4]^* & =  1/3 \ ,\quad\\
    \left[q_2,q_3\right]^* = [q_2,p_2]^* = -[q_2,p_4]^* & =  1/3 \ ,\\
    \left[q_3,q_4\right]^* = -[q_4,p_2]^* = [q_4,p_4]^* & =  2/3 \ ,
\end{align}\ese
with all other brackets~vanishing. 

The constraints can be solved in various ways and the results can be used freely, either before or after computing Dirac brackets. For~example, the~constraints imply
\bse\label{justq4andp2}\begin{align}
    q_1 & =  -q_4/2 \ ,\\
    q_2 & =  -q_4/2 \ ,\\
    q_3 & =  p_2 \ ,\\
    p_1 & =  0 \ ,\\
    p_3 & =  0 \ ,\\
    p_4 & =  -p_2 \ .
\end{align}\ese
We can use these to eliminate the variables $q_1$, $q_2$, $q_3$, $p_1$, $p_3$ and $p_4$. 
Then 
 the extended Hamiltonian $H_E$ becomes the {\em fully reduced Hamiltonian} 
\be
    H_{FR} = \frac{1}{2}\left[ p_2^2 + \frac{9}{4} q_4^2\right]  \ ,
\ee
which depends only on $q_4$ and $p_2$. 

The Dirac brackets of the variables that remain are $[q_4,p_2]^* = -2/3$. Thus, the~equations of motion become
\bse\label{HamsEqnsHFR}\begin{align}
    \dot q_4 = [q_4,H_{FR}]^* & =  -2p_2/3  \ ,\\
    \dot p_2 = [p_4,H_{FR}]^* & =  3q_4/2 \ .
\end{align}\ese
These are the equations for a simple harmonic oscillator with solution  
\bse\begin{align}
    q_4(t) & =  \alpha\sin t + \beta\cos t \ ,\\
    p_2(t) & =  -(3\alpha/2)\cos t + (3\beta/2)\sin t \ ,\quad
\end{align}\ese
where $\alpha$ and $\beta$ are arbitrary~constants. 

With the gauge fixed,
 the dynamics take place on the fully reduced phase space, the~two-dimensional surface defined by the constraints and gauge conditions ${\cal C}^{(all)}_A = 0$. There are many different choices of coordinates for this surface. Instead of solving the constraints for $q_4$ and $p_2$, we could solve them for $q_2$ and $q_3$. 
 In~that case,
  the fully reduced Hamiltonian is  
\be
    H_{FR} = \frac{9}{2} q_2^2 + \frac{1}{2} q_3^2
\ee
and Hamilton's equations are 
\bse\begin{align}
    \dot q_2 &  = [q_2,H_{FR}]^* = q_3/3 \ ,\\
    \dot q_3 & = [q_3,H_{FR}]^* = -3 q_2 \ 
\end{align}\ese
Again, this describes the simple harmonic~oscillator. 

We can choose other coordinates on the fully reduced phase space.  For~example, let $q_4 = Q + q_2$ and $p_2 = -P$, then use the constraints to eliminate $q_1$, $q_2$, $q_3$, $p_1$, $p_3$ and $p_4$. The~fully reduced Hamiltonian becomes 
\be
    H_{FR} = \frac{1}{2} ( Q^2 + P^2 ) \ .
\ee
The nonzero Dirac brackets are $[Q,P]^* = 1$, and~the equations of motion are simply $\dot Q = P$ and $\dot P = -Q$. 

In each case, the~fully reduced theory exhibits the single physical degree of freedom that we~expect. 

\section{Fully Reduced~Action}\label{sec:fullyreducedS}
The fully reduced equations of motion can be derived from the action that includes all of the constraints and gauge conditions. For~lack of a better name, let us denote this action with the subscript ``all'': 
\be
    S_{all} = \int_0^T dt \left\{ p_i \dot q_i - H_{fc} - \gamma_a {\cal C}_a^{(fc)}  
    - \sigma_a{\cal C}_a^{(sc)} - \rho_a {\cal G}_a  \right\} \ .
\ee
Now 
 extremize $S_{all}$ with respect to variations in $p_1$, $p_3$, $p_4$, $q_1$, $q_2$, $q_3$ and $q_4$: 
\bse\label{someqpeqns}\begin{align}
    \dot q_1 & =  \gamma_1 + \sigma_1/3 \ ,\\
    \dot q_3 & =  -q_1 + p_3 + p_4/2 + \gamma_1/2 + \gamma_2  - \sigma_1/3 + \sigma_2/3 \ ,\\
    \dot q_4 & =  -p_2/2 + p_3/2 - \gamma_1/2 + \sigma_1/3 + \rho_2 \ ,\\
    \dot p_1 & =  q_1 + q_2 + q_4 + p_3 - \sigma_2 - \rho_1 \ ,\\
    \dot p_2 & =  q_1 + 2q_4 + \sigma_2 - \rho_1 \ ,\\
    \dot p_3 & =  -\rho_2 \ .
\end{align}\ese
In addition,  vary $S_{all}$ with respect to the Lagrange multipliers to obtain the constraints ${\cal C}^{(all)}_A = 0$. The~solution of the full set of equations, (\ref{someqpeqns}) and ${\cal C}^{(all)}_A = 0$, is given by Equations~(\ref{justq4andp2}) along with 
\bse\begin{align}
    \gamma_1 & =  -p_2/3 - 2\dot p_3/2 + 2\dot q_1/3 - 2\dot q_4/3 \ ,\qquad\\
    \gamma_2 & =  p_2 - 3q_4/4 + \dot p_1/6 + \dot p_2/6  + \dot p_3 + \dot q_3 + \dot q_4 \ ,\\
    \sigma_1 & =  p_2 + 2\dot p_3 + \dot q_1 + 2\dot q_4 \ ,\\
    \sigma_2 & =  3q_4/4 - \dot p_1/2 - \dot p_2/2 \ ,\\
    \rho_1 & =  -3q_4/4 - \dot p_1/2 + \dot p_2/2 \ ,\\
    \rho_2 & =  -\dot p_3 \ .
\end{align}\ese
We now insert these results into $S_{all}$ to obtain 
the {\em fully reduced action}
\be
    S_{FR}[q_4,p_2] = \int_0^T dt \left\{  -\frac{3}{2} p_2\dot q_4 -  \frac{1}{2}\left[ p_2^2 + \frac{9}{4} q_4^2 \right] \right\} \ ,
\ee
which is a functional of $q_4$ and $p_2$. 
The equations of motion $\delta S_{FR} = 0$ are 
\bse\begin{align}
   0 & =   \frac{\delta S_{FR}}{\delta q_4} = \frac{3}{2} \dot p_2 - \frac{9}{4} q_4  \ ,\\
   0 & =  \frac{\delta S_{FR}}{\delta p_2} = -\frac{3}{2} \dot q_4 - p_2 \ .
\end{align}\ese
These are equivalent to 
Hamilton's Equations~(\ref{HamsEqnsHFR}) for the fully reduced~Hamiltonian. 

We can place $S_{FR}$ into ``canonical form'' by defining new variables $P = -p_2$ and $Q = 3q_4/2$. 
Then 
\be
    S_{FR}[Q,P] = \int_0^T dt \left\{ P\dot Q - \frac{1}{2}[P^2 + Q^2] \right\} \ ,
\ee
which is the familiar action for the harmonic~oscillator. 
    
\section{Summary and~Discussion}\label{sec:summary}
Here is the Dirac--Bergmann~algorithm: 
\begin{itemize}
\item Compute the conjugate momenta $p_i = \partial L/\partial\dot q_i$ and define the canonical Hamiltonian $H_C$ as $p_i \dot q^i - L(q,\dot q)$, written in terms of $p$'s and $q$'s.
\item Identify the primary constraints. The~primary Hamiltonian $H_P$ is obtained from $H_C$ by adding the primary constraints with Lagrange multipliers. 
\item Apply Dirac's consistency conditions to identify higher-order constraints and restrictions on the Lagrange multipliers. 
\item The total Hamiltonian $H_T$ is found from $H_P$ by incorporating the restrictions on the Lagrange 
multipliers. 
\item Separate the primary, secondary, and~higher-order constraints into first and second class.
\item The first class Hamiltonian $H_{fc}$ is the part of $H_T$ with the primary first class constraints removed. 
\item The extended Hamiltonian $H_E$ is obtained from the first class Hamiltonian $H_{fc}$ by adding all of the first class constraints with Lagrange multipliers. 
\item The partially reduced Hamiltonian $H_{PR}$ is found from $H_E$ by using Dirac brackets to eliminate the second class constraints.
\item Gauge freedom is removed by assigning gauge conditions. 
The fully reduced Hamiltonian $H_{FR}$ is obtained from $H_E$ by using Dirac brackets to impose all constraints and gauge conditions. 
\end{itemize}
The theory defined by the singular Lagrangian (\ref{TheLagrangian}) provides a relatively complete example of each step in the~algorithm. 

One reason the Dirac--Bergmann algorithm can be confusing is that typical examples are chosen for simplicity, allowing some of the 
logical steps to be skipped. This causes the distinction between Hamiltonians to become blurred. 
For example, if~there are no restrictions on the Lagrange multipliers, then the primary Hamiltonian $H_P$ and the total Hamiltonian $H_T$ coincide. Likewise, if~there are no secondary (or higher-order) first class constraints, then the total Hamiltonian $H_T$ and the extended Hamiltonian $H_E$ coincide. Moreover, for~theories with no second class constraints and no gauge conditions imposed, Dirac brackets and the reduction process are not~needed. 

Another confusing aspect of the Dirac--Bergmann algorithm 
is that for many important theories, the~Lagrangian is given to us in a  form  that contains Lagrange multipliers. For~example, consider the Einstein--Hilbert action of general relativity. The~Lagrangian density is the spacetime curvature scalar. A~3+1 splitting of the spacetime metric~\cite{MTW,Arnowitt:1962hi} allows us to write the Lagrangian density as
\be\label{EHLagDens}
    L = R + K_{ij}(g^{ik}g^{j\ell} - g^{ij}g^{k\ell} )K_{k\ell} \ ,
\ee
apart from a total derivative term that integrates to the boundary. 
Here, $R$ and $g_{ij}$ are the spatial scalar curvature and spatial metric.  
In addition, 
$K_{ij}$ is the extrinsic curvature of space, built from the lapse function $N$, shift vector $N^i$ and spatial  derivatives of $g_{ij}$. 
The Lagrangian density depends on the Lagrange multipliers $N$ and $N^i$ as well as the configuration space coordinates $g_{ij}$.

Because the Einstein--Hilbert Lagrangian (\ref{EHLagDens}) depends on Lagrange multipliers, it is not analogous to the singular Lagrangian (\ref{TheLagrangian}) of our example problem. Rather, it is analogous to the partially reduced
Lagrangian that appears in the integrand of the action $S_{PR}[Q,\gamma]$ of Equation~(\ref{SPRofQgamma}).  Recall that the partially reduced Hamiltonian $H_{PR}$ contains only first class constraints and no restrictions on the Lagrange multipliers. Likewise, the~Hamiltonian for general relativity~\cite{Arnowitt:1962hi} is constructed from first class constraints (the Hamiltonian and momentum constraints), and~the Lagrange multipliers (the lapse function and shift vector) are~unrestricted. 

We can attempt to eliminate the lapse and shift from the Einstein--Hilbert action, just as we eliminated the $\gamma$'s from $S_{PR}[Q,\gamma]$ and obtained the result $S_{PR}[Q]$ in Equation~(\ref{PRactionofQ}). 
It is  straightforward to eliminate the lapse $N$---the result is the Baierlein--Sharp--Wheeler action~\cite{BaierleinSharpWheeler}. However, the~shift vector cannot be eliminated algebraically because the equations of motion obtained by varying the action with respect to the lapse and shift depend on  spatial derivatives of $N^i$. 

So,
 for  general relativity,
  we do not have a singular Lagrangian analogous to Equation~(\ref{TheLagrangian}), and~we cannot expect to apply the Dirac--Bergmann 
algorithm from beginning to end  as laid out by Dirac~\cite{DiracLectures}. 
Nevertheless, the~general Dirac--Bergmann algorithm serves as the foundation  for our understanding and interpretation of the Hamiltonian form of the~theory.

\begin{acknowledgments}
I would like to thank Claudio Bunster and Marc Henneaux for sparking my interest in this subject many years ago. I also thank Ashiqul Islam Dip for helpful discussions. 
\end{acknowledgments}


\bibliography{references}
\bibliographystyle{unsrt}

\end{document}